\documentclass[aps,prb,final,twocolumn,lengthcheck,superscriptaddress,showpacs,amsmath,amssymb]{revtex4}

\usepackage{graphicx}
\usepackage{dcolumn}
\usepackage{bm}
\usepackage[T1]{fontenc}
\usepackage{ae,aecompl}

\begin{document}


\title{Topology-induced confined superfluidity in inhomogeneous arrays}

\author{P. Buonsante}
\affiliation{Dipartimento di Fisica, Politecnico di Torino and I.N.F.M, Corso Duca degli Abruzzi 24 - I-10129 Torino (ITALIA)}%
\author{R. Burioni}
\affiliation{Dipartimento di Fisica, Universit\`a degli Studi di Parma and I.N.F.M., Parco Area delle Scienze 7/a I-43100 Parma (ITALIA)}%
\author{D. Cassi}
\affiliation{Dipartimento di Fisica, Universit\`a degli Studi di Parma and I.N.F.M., Parco Area delle Scienze 7/a I-43100 Parma (ITALIA)}%
\author{V. Penna}%
\affiliation{Dipartimento di Fisica, Politecnico di Torino and I.N.F.M, Corso Duca degli Abruzzi 24 - I-10129 Torino (ITALIA)}%
\author{A. Vezzani}
\affiliation{Dipartimento di Fisica, Universit\`a degli Studi di Parma and I.N.F.M., Parco Area delle Scienze 7/a I-43100 Parma (ITALIA)}%

\date{\today}

\begin{abstract}
We report the first study of the zero-temperature phase diagram 
of the Bose-Hubbard model
on topologically inhomogeneous arrays.  We show that the 
usual Mott-insulator and 
superfluid domains, in the paradigmatic case
of the comb lattice, are separated by regions where the superfluid 
behaviour of the bosonic system is confined along the comb backbone.
The existence of such {\it confined superfluidity}, arising from topological 
inhomogeneity, is proved by different analytical and numerical techniques 
which we extend to the case of inhomogeneous arrays. 
We also discuss the relevance of our results to real system exhibiting 
macroscopic phase coherence, such as coupled Bose condensates and 
Josephson arrays.
\end{abstract}

\maketitle

\section{Introduction}
The Bose-Hubbard (BH) Hamiltonian, describing bosons hopping across the sites of a discrete structure and originally introduced to model liquid He in confined geometries, \cite{A:Fisher} proves successful in capturing the essential physics of a wide range of condensed-matter systems. 
The best known examples 
are no doubt provided by Josephson-junction arrays
(JJA)\cite{A:FazioPR,A:Kuo} and Bose-Einstein condensate (BEC) arrays,
\cite{A:Jaksch,A:Greiner02} which are the subject of
a huge amount of both theoretical and experimental ongoing investigations. 
The hallmark of such class of systems is the
presence of a superfluid phase as opposed to a (Mott) 
insulator phase.\cite{B:Sachdev}
The theoretical studies hitherto carried out on such phase transition mostly 
focus on homogeneous ambient  lattices, 
and provide well-established numerical and analytical techniques. Homogeneous lattices are also the basis of the current experimental
realizations of systems belonging to the BH class. This is at least
partly due to present technical constraints. For instance the optical
techniques used to fragment BECs  yield quite naturally homogeneous 
arrays.  

However, the striking progress in experimental techniques suggests the 
realization of inhomogeneous networks to be at hand. 
Actually, JJAs can be engineered in nontrivial 
geometries with the only possible
constraint of planarity. In this respect, interesting geometry-driven effects
are proposed in Refs. \onlinecite{A:Doucot,A:Giusiano}, 
while the physics of a fractal JJA is 
experimentally studied in Ref. \onlinecite{A:Meyer02}.
As to BEC arrays, two very promising 
approaches for realizing inhomogeneous topologies are provided  by
holographic optical traps \cite{A:Pu,A:Grier} and magnetic microtraps \cite{A:Folman01,A:Ott,A:Hansel}. In the latter case, ongoing efforts are aimed at reducing the
spacing between individual microtraps, currently bound above a few $\mu$m, 
in order to couple the condensates therein confined. 

The deep influence of topological inhomogeneities on thermodynamic properties
of discrete boson systems is evidenced by the occurrence of unexpected
features even in the absence of boson interaction.
Indeed, a finite-temperature
Bose-Einstein condensation can take place despite the low 
dimensionality of the system.
This is illustrated in Refs. \onlinecite{A:CombL,A:MesoComb}
in the 
case of the square comb lattice, 
namely an array of linear chains
(fingers) joined along a transverse direction (backbone) such as the one shown
in the inset of Fig. \ref{F:MFpd}.
More precisely, inhomogeneity induces a {\it hidden} band in the single-particle 
energy spectrum which  is ultimately responsible for condensation.

On the other hand, the rich zero-temperature phase diagram of the BH model
ensues from the competition between the on-site 
repulsion and the kinetic energy
of the boson gas. In the light of this, a natural question arises as to the
influence of topology on the physics of interacting bosons.
In this respect we mention that the effect of the inhomogeneity arising from the superposition of a local 
on-site potential on an otherwise regular lattice has been recently addressed.
In particular, the existence of local Mott domains induced by a parabolic confining 
potential has been evidenced for BEC arrays in Refs. \onlinecite{A:Polkov,A:Kashurnikov02,A:Batrouni02,A:Pupillo03,A:Rey}
 while Refs. \onlinecite{A:Roth03,A:Aizenman,A:Santos,A:LobiMF} analyze 
the phase diagram on superlattices.

Here, we consider inhomogeneities of purely topological (i.e. kinetic rather than potential) origin, 
focusing on the emblematic case of comb lattices, where the larger connectivity of the backbone is 
expected to act as a catalyst for superfluidity. Interestingly, the competition between 
kinetic and boson-interaction energy causes the occurrence of an intermediate domain in the BH phase diagram. 
The  usual Mott-insulator  lobes are separated  from the  superfluid domain  by a
phase characterized  by the localization of superfluidity  in a narrow
region  surrounding  the comb  backbone,  the  rest  of the  structure
exhibiting an  unexpected insulator-like behavior.  More precisely, we  show that
the  {\it  local   compressibility}  \cite{A:Batrouni02}  features  an
exponential decrease with increasing  distance from the backbone. Note
indeed that the topological  inhomogeneity of the structure requires a
description in  terms of site-dependent  quantities.
These results, which --- to the best of our knowledge --- 
are the first concerning the influence of topology on the BH phase diagram, 
required the generalization of different numerical and analytical techniques \cite{A:scpe}. 
The presence of {\it confined superfluidity} is first  evidenced within 
a mean-field  approach, and further 
confirmed by both a third order analytical strong coupling perturbative expansion
(SCPE) and (population) quantum Monte-Carlo (QMC) simulations.     
  
\section{The Bose-Hubbard model on a generic structure}
The BH Hamiltonian, describing locally interacting bosons on a generic discrete structure consisting of $M$ sites, is
\begin{equation}
\label{E:BH}
H \!= \!\sum_{j=1}^{M}\left[\frac{U}{2} n_j (n_j-1)\!-\!\mu n_j\right] -T\sum_{h,j=1}^{M} A_{h\, j} a_h a_j^+ 
\end{equation}
where the operator $a_j^+$ ($a_j$) creates (annihilates) a boson at site $j$ and $n_j=a_j^+ a_j$ counts the bosons sitting at site $j$. As to the parameters, $U>0$ accounts for the (on-site) repulsion among bosons, $\mu$ is the chemical potential and $T$ is the hopping amplitude between adjacent sites, specified by the adjacency matrix $A$. 
This is a useful tool supplied by graph theory,\cite{B:Harari} allowing an algebraic description of the topology of a generic discrete structure. Its  generic matrix element $A_{h\,j}$ is one if sites  $(h,j)$ are nearest neighbors and zero otherwise. 
In view of  $[N,H]=0$, where $N=\sum_j n_j$,  Hamiltonian (\ref{E:BH}) can be conveniently studied exploiting 
 its block-diagonal structure. 
Since we are interested in the zero temperature phase diagram, 
hereafter  $\langle \cdot \rangle$ denotes the expectation value on the ground state of $H$.

As we mentioned above, in the case of homogeneous topology the competition between  on-site interaction and  hopping  gives rise to an interesting zero-temperature phase-diagram in the  $\mu/U$-$T/U$ plane, where two different domains can be recognized. A superfluid phase, where the energy cost of adding or subtracting a boson to the system vanishes in the thermodynamic limit; An incompressible Mott-insulator phase, 
consisting of a series of adjacent lobes, 
where such operations cost a finite amount of energy and the filling $f\equiv N/M$ is pinned to an integer value. 
The Mott-superfluid transition can be furthermore characterized by the compressibility $\kappa = \partial N/\partial \mu$, which is finite in the superfluid region and vanishes within the Mott lobes. In the case of inhomogeneous systems, the possible effects of topology can be described in detail by the site-dependent {\it local compressibility},\cite{A:Batrouni02} $\kappa_j = \partial \rho_j/\partial \mu$, where $\rho_j = \langle n_j \rangle$ is the local density of bosons.

Owing to the enormous size of the Fock space, an exact solution of the model cannot be faced even for relatively small structures. However, the essential elements of the Mott-superfluid transition can be captured resorting to different approximate schemes, such as mean-field,\cite{A:Sheshadri,A:Amico2}
 SCPE,\cite{A:Freericks1} the renormalization approach \cite{A:Kuehner} and QMC computations.\cite{A:Batrouni,A:Kashurnikov96}

\section{Mean-field approximation}

The key point of the mean-field approach of Ref.~\onlinecite{A:Sheshadri} consists in the approximation 
\begin{equation}
\label{E:mfa}
\left(a_h -\langle a_h\rangle \right) \left(a_j^+ -\langle a_j^+\rangle \right)  \approx 0
\end{equation}
allowing to recast Hamiltonian (\ref{E:BH}) as the sum of on-site Hamiltonians $H \approx {\cal H} = \sum_j {\cal H}_j$. 
In the simple case of a $d$-dimensional (translationally invariant) lattice, ${\cal H}_j$ is site independent and
one is left with a single site problem \cite{A:Sheshadri} 
\[
{\cal H}(\alpha) = M [\frac{U}{2} n(n-1)-\mu n -2 d\, T(a+a^+) \alpha + 2 d\, T \alpha^2],
\]
 subject to the self consistency constraint $\alpha = \langle a \rangle$,\cite{N:Number} where the so-called {\it superfluid parameter} $\alpha$ can be considered real without loss of generality.
 The phase diagram of the homogeneous case can be obtained numerically \cite{A:Sheshadri}  or even analytically. \cite{A:vanOosten,A:LobiMF} 

For non-homogeneous structures
\begin{eqnarray}
\label{E:mfBH}
{\cal H}\left(\{\alpha_h\}\right) &=& \sum_{j=1}^M {\cal H}_j\\
{\cal H}_j &=& \frac{U}{2} n_j (n_j-1)-\mu n_j \nonumber\\ 
&-& T\! \sum_{h=1}^{M} A_{j h} \alpha_h \left( a_j + a_j^+ - \alpha_j \right) 
\end{eqnarray}
and the ground state of ${\cal H}$ has the form  $|\psi\rangle = \bigotimes_j |j; \{\alpha_h\}\rangle $, where $ |j; \{\alpha_h\}\rangle $ is the ground state of ${\cal H}_j$. Thus the problem is solved by finding the set of real quantities $\{\alpha_h\}_{h=1}^M$ such that 
\begin{equation}
\label{E:sce}
\langle j;\{\alpha_h\}|a_j|j;\{\alpha_h\}\rangle = \alpha_j.
\end{equation}
 This can be easily done numerically by means of  self-consistent iterative algorithm.\cite{A:Polkov,A:LobiMF} 
 Despite the approximation in Eq.~(\ref{E:mfa}) strongly suppresses spatial correlation, some topological information is retained in the above mean-field formulation owing to the presence of the adjacency matrix $A$ in Hamiltonian~(\ref{E:mfBH}). In this case $\rho_j$, $k_j$ and $\alpha_j$ are in general site-dependent quantities.
 For comb lattices these quantities are constant along the backbone direction, owing to the symmetry of the system.
 
\begin{figure}
\begin{center}
\includegraphics[width=8.5cm]{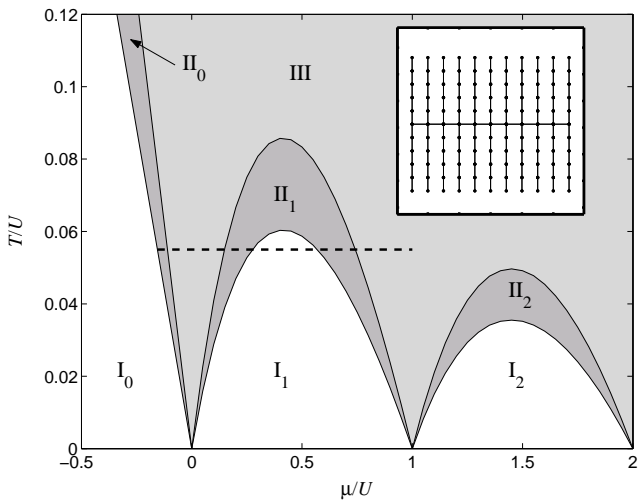}
\caption{\label{F:MFpd} {\bf Inset:} an example of square 
comb lattice featuring 11 sites both on the backbone and on the ribs \cite{N:per}; {\bf Main plot}: Mean-field Phase diagram of a 100$\times$100 comb lattice. Different shades of grey denote different phases. The dashed line at $T/U = 0.055$ signals the set of parameters considered in Fig.~\ref{F:MFdns}. }
\end{center}
\end{figure}
\begin{figure}
\begin{center}
  \includegraphics[width=8.5cm]{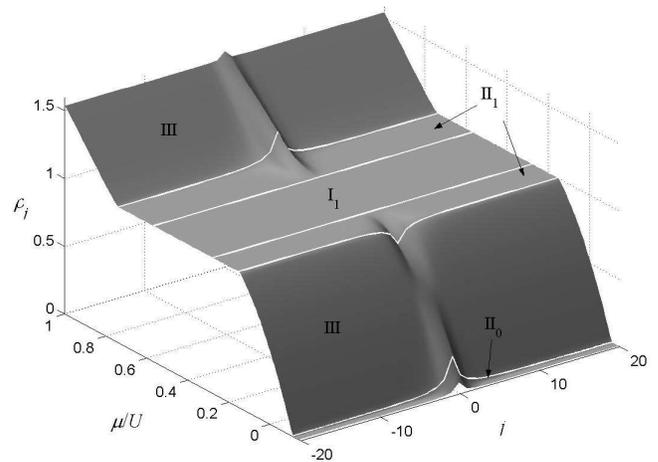}
\caption{\label{F:MFdns} Behaviour of the local density of bosons $\rho_j$ for sites $j$ along a finger of the comb lattice ($j=0$ backbone). The figure refers to a fixed value of the hopping amplitude $T/U=0.055$ and a finite interval of $\mu/U$ (dashed line in Fig.~\ref{F:MFpd}). The white density profiles correspond to the borders of the different regions of the phase diagram. As we discussed in the text $\rho_j=1$ inside region I$_1$, $\rho_j \to f$ in regions II$_f$, whereas $\rho_j$ tends to a not necessarily integer number in region III.   }
\end{center}
\end{figure}

Fig.~\ref{F:MFpd} shows the numerically determined mean-field phase diagram for a $100\times100$ comb lattice. In the regions ${\rm I}_f$ (where $f$  is the integer filling)  $\kappa_j =  0$ for all $j$'s and the total number of bosons is pinned at $N=f M$. The system is therefore an incompressible Mott insulator. In the  regions ${\rm II}_f$, $\kappa_j$ is finite, yet it vanishes exponentially along the fingers. The same behaviour is observed for $\rho_j$, which  is exponentially close to $f$ with increasing distance from the backbone. Hence in these regions the superfluid behaviour of the system is confined along the backbone direction alone. An extended  superfluid behaviour is recovered in region III, where the local density $\rho_j$ far from the backbone is not necessarily an integer quantity, and $\kappa_j$ is nowhere vanishing. Such behaviour of the local density of bosons and compressibility are summarized in Fig.~\ref{F:MFdns} and in the upper panel of Fig.~\ref{F:lc}, respectively. 

We mention that it is possible to evaluate the exact analytical form of the boundaries of the I$_f$ regions as provided by the mean-field approach described by Eqs.~(\ref{E:mfBH})--(\ref{E:sce}). This can be for instance accomplished making use of the finite-temperature method reported in Ref.~\onlinecite{A:LobiMF}, and subsequently letting the temperature go to zero. The function of $T/U$ describing the I$_f$ boundary for a generic structure characterized by the adjacency matrix $A$ is obtained by rescaling the corresponding function for a homogeneous $d$-dimensional lattice \cite{A:vanOosten,A:LobiMF} by a factor $\frac{\lambda}{2 d}$, where $\lambda$ is the maximal eigenvalue of $A$.

In general,
topological inhomogeneities make the study of critical behaviours a rather difficult task.
However, the above results suggest some considerations  in this respect. 
On regular lattices the correlation length diverges in any direction at a critical point. 
Conversely, on the comb lattice, the correlation length is expected to diverge only along a specific direction, 
depending on the critical border under concern. 
To wit, the correlation length between sites of the same finger is   
the only divergent quantity at the  
II$_f$--III transition, while it is finite at the border between regions I$_f$ and II$_f$ (where the divergent quantity
is the correlation length between sites of different fingers). 
We also mention that preliminary results based on the mean-field approach of Ref.~\onlinecite{A:LobiMF} 
indicate that the above picture is robust at small finite temperatures, and hence in principle 
accessible to experiments. In this respect we note that the three different phases in 
Fig.~\ref{F:MFpd} can be probed as in Ref.~\onlinecite{A:Greiner02}, provided that the system is 
realized in terms of coupled BECs, possibly using holographic traps.\cite{A:Pu,A:Grier} 
Indeed, after the trapping potential is released, the expanding atomic clouds should produce either a one-dimensional\cite{A:Pedri} or 
a two-dimensional\cite{A:Adhikari} interference pattern depending on whether superfluidity is confined along the 
backbone or extended on the entire comb. 

\section{Beyond mean-field}

\begin{figure}
\begin{center}
\includegraphics[width=8.7cm,angle=0]{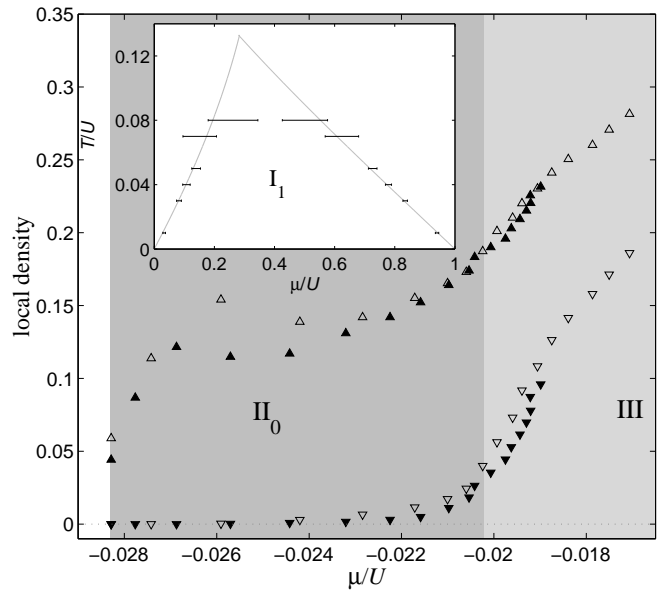}
\caption{\label{F:buckle} {\bf Inset}: Mott lobe I$_1$ for a comb lattice according to third order 
SCPE (solid line) and population QMC (errorbars). 
The latter refer to a 12$\times$12 lattice. 
{\bf Main plot}: QMC results for the on-site density of bosons on a comb lattice at $T/U = 0.01$. 
Upward triangles: average result for a site on the backbone. 
Downward triangles: average result for the farthest site from the backbone. 
Open and filled symbols refer to a 12$\times$12 and a 16$\times$16 comb lattice, respectively. 
The larger error on the QMC data (abscissae) is smaller than the symbol size. }
\end{center}
\end{figure}

A step beyond the mean-field approximation consists in the strong coupling perturbative expansion. Indeed,  time-independent perturbation theory in the hopping parameter allows to obtain an analytical approximation of the Mott lobes, which, in the case of the linear chain, proves to be quite satisfactory already at the third order.\cite{A:Freericks1} This approach, introduced in Ref.~\onlinecite{A:Freericks1} for homogeneous bipartite structures, is extended to any structure in Ref.~\onlinecite{A:scpe}. Quite interestingly, it turns out that topological inhomogeneity gives rise to a third-order correction featuring an unusual dependence on the adjacency matrix describing the topology of the structure. Indeed, unlike the previously reported perturbative terms, depending only on the maximal eigenpair of $A$, the ``topological correction'' depends on the entire spectrum of the adjacency matrix. The solid line in the inset of Fig.~\ref{F:buckle} is the border of the Mott lobe I$_1$ for a comb lattice as provided by the analytical third-order strong-coupling perturbative expansion reported in Ref.~\onlinecite{A:scpe}. The above described exponential localization  of superfluidity  characterizing phases II$_f$ is also captured by SCPE even at  order zero. 
Indeed it can be easily shown that $\rho_j = f + C |v_j|^2$ and $\kappa_j = K  |v_j|^2$, where $C$ and $K$ are normalizing constants and $v_j$ is the $j$-th component of the maximal eigenvector of the adjacency matrix, depending only on the distance $d_j$ from the backbone,\cite{A:CombL,A:MesoComb} $v_j =\exp[-d_j  {\rm asinh} (1)]$. First order SCPE confirms this behaviour, though with a different decay rate, as well as the above considerations about the correlation functions.\cite{A:scpe}
\begin{figure}
\begin{center}
\includegraphics[width=8.5cm,angle=0]{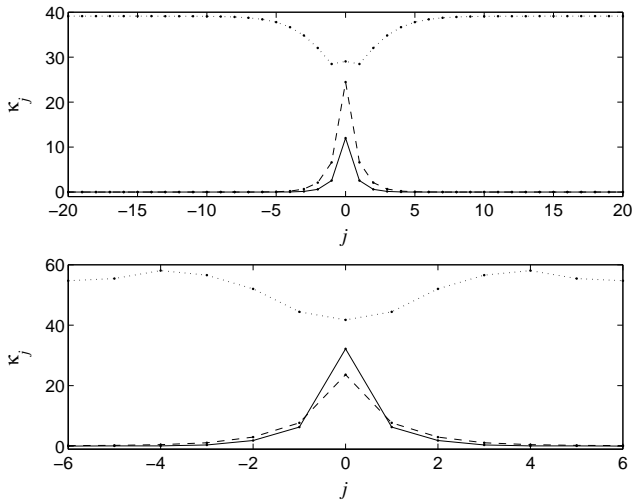}
\caption{\label{F:lc} Local compressibility $\kappa_j$ for sites $j$ along one finger of the comb lattice ($j=0$ is on the backbone). Dashed, solid and dotted lines refer to region II$_0$, II$_1$ and III, respectively. Note the exponential decrease with increasing distance from the backbone characterizing the first two curves. {\bf Upper panel}: Mean-field result. {\bf Lower panel}: QMC data for a 12$\times$12 lattice. }
\end{center}
\end{figure}


The data points in the inset of Fig.~\ref{F:buckle}, satisfactorily agreeing with the perturbative curve, have been obtained with a  population QMC approach \cite{A:Iba} adopting the resampling procedure described in Ref.~\onlinecite{A:Nightingale}. Based on a generalization of the power method for finding the maximal eigenpair of a matrix, this technique essentially amounts to a stocastic evaluation of the ground state of Hamiltonian (\ref{E:BH}) and therefore allows to study the zero temperature phase diagram of the BH model.
Note that both in the SCPE  and in the QMC approach $\mu$ is evaluated as the difference between the ground state energies of systems whose total number of bosons differs by one.\cite{A:Batrouni,A:Freericks1} In particular, the border of phases I$_f$ are obtained considering the energy cost of adding or subtracting one boson from the integer filling situation $N= M f$.

The QMC approach also confirms the existence of the confined superfluid phases II$_f$. Fig.~\ref{F:buckle} clearly shows the transition between phases II$_0$ and III. Indeed in the former region the local density $\rho_j$ at a site far from the backbone (downward triangles) is very close to $f=0$, and it  features a sudden increase only after entering phase III. Conversely, the local density on the backbone (upward triangles) is fractional also in region II$_0$, thus confirming that the superfluid is localized there. 

A further confirmation of confined superfluidity is provided by the local compressibility profiles appearing in the lower panel of Fig.~\ref{F:lc}, obtained by means of QMC simulations in the case of a 12$\times$12 comb lattice. Indeed, in the extended superfluid region the local compressibility is everywhere significantly larger than zero, whereas the curves relevant to the regions II$_0$ and II$_1$ feature a sharp decrease with increasing distance from the backbone.
Note that the local compressibility within the Mott lobes I$_f$ is everywhere zero, since, by definition, the ground state of the system can be changed only if the chemical potential $\mu$ is varied of an amount sufficient to cross the lobe border.

\section{Conclusions}

In summary, we reported the first analysis of the influence of topological
inhomogeneity on the phase diagram of interacting bosons, considering the
emblematic case of comb lattices. This supplies a basis and a conceptual
framework for a more general study aimed  at a deeper understanding of the
role of topology in quantum phase transitions. Furthermore we suggested a
possible experimental setup, based on BEC arrays trapped in holographic
potentials,\cite{A:Pu,A:Grier} where the intermediate phase occurring on a 
comb lattice could be observed.\cite{A:Greiner02,A:Pedri,A:Adhikari}

The recently disclosed relation
between critical behaviour and system-state entanglement \cite{A:Osterloh}
provides a further context where the influence of geometry might play a
significant role. In particular, inhomogeneous arrays have been recently
proposed as quantum-information-processing devices.\cite{A:Giorda03}  We point
out that, owing to the formal mathematical analogy between Heisenberg and BH
models (see e.g. Refs. \onlinecite{A:FazioPR} and \onlinecite{A:Rojas}),  the
results herewith presented have also relevant implications for quantum spin
systems on inhomogeneous structures. As a concluding remark, we observe that a
comb lattice can be obtained by joining 1D structures or appropriately removing
the exceeding links from a 2D regular array.\cite{A:Giusiano} This makes the
structures considered here  not only interesting from the theoretical
point of view, but also very promising for actual realizations based on
JJA technology.\cite{A:Kuo,A:Meyer02,A:Yamaguchi}

\acknowledgments
The work of P.B. and A.V.
has been supported by MURST project {\it Quantum Information and Quantum
Computation on Discrete Inhomogeneous Bosonic Systems}. 


\begin{thebibliography}{10}
\bibitem{N:Number}
Note that, unlike {\ref{E:BH}}, the MF Hamiltonian $\cal H$ does not conserve
  the number so that $\langle a\rangle = \alpha$ is not necessarily zero.
\end{thebibliography}

\end{document}